\documentclass[fleqn,usenatbib]{mnras}
\usepackage{newtxtext,newtxmath}

\usepackage[T1]{fontenc}
\usepackage{ae,aecompl}

\usepackage{graphicx}	
\usepackage{amsmath}	
\usepackage{amssymb}	
\usepackage{pdflscape}
\usepackage{soul} 

\newcommand{\x}{\,$\times$}
\newcommand{\krange}[2]{${#1}\leq~k<{#2}$}

\title[Outlier Scoring]{Density Based Outlier Scoring on Kepler Data}

\author[D. Giles and L. Walkowicz]{
Daniel K. Giles,$^{1,2}$\thanks{E-mail: dgiles1@hawk.iit.edu}
Lucianne Walkowicz,$^{1}$\thanks{E-mail: lwalkowicz@adlerplanetarium.org}
\\
$^{1}$Astronomy Department, The Adler Planetarium, Chicago, IL 60605\\
$^{2}$Physics Department, Illinois Institute of Technology, Chicago, IL, 60616\\
}

\date{Accepted XXX. Received YYY; in original form ZZZ}

\pubyear{2020}

\begin{document}
\label{firstpage}
\pagerange{\pageref{firstpage}--\pageref{lastpage}}
\maketitle

\begin{abstract}
In the present era of large scale surveys, big data presents new challenges to the discovery process for anomalous data. Such data can be indicative of systematic errors, extreme (or rare) forms of known phenomena, or most interestingly, truly novel phenomena which exhibit as-of-yet unobserved behaviors. In this work we present an outlier scoring methodology to identify and characterize the most promising unusual sources to facilitate discoveries of such anomalous data. We have developed a data mining method based on k-Nearest Neighbor distance in feature space to efficiently identify the most anomalous lightcurves. We test variations of this method including using principal components of the feature space, removing select features, the effect of the choice of k, and scoring to subset samples. We evaluate the peformance of our scoring on known object classes and find that our scoring consistently scores rare (<1000) object classes higher than common classes. We have applied scoring to all long cadence lightcurves of quarters 1 to 17 of Kepler's prime mission and present outlier scores for all 2.8 million lightcurves for the roughly 200k objects.
\end{abstract}



\section{Introduction}
\label{sec:intro}
Modern astronomical surveys are looking deeper into space, covering greater regions of the sky, and taking more data for more sources than ever before.
The unprecedented scale of modern surveys has presented significant data mining challenges, with science goals ranging from discovering and studying potentially habitable exoplanets, expanding the databases of every known phenomena, to discovering entirely new signals from sources which have yet to be theorized or observed.
NASA's Kepler Mission surveyed \textasciitilde 200,000 sources in a single region of space for over four years, producing over 2 million lightcurves in its primary mission, and the Transiting Exoplanet Survey Satellite \citep[TESS;][]{TESS} is about midway through it's primary mission to survey the entire sky, both in the search for habitable exoplanets. Gaia has released data for close to 2 billion sources \citep{Gaia2016,Gaia2018} generating the most comprehensive catalog of astronomical objects to date and growing, and the Zwicky Transient Facility \citep[ZTF;][]{ZTF} has released data consisting of millions of images and over a billion lightcurves using an extremely wide field and fast readout to identify new and rare transients. 
The Vera Rubin Observatory is slated to have first light in the coming years and will deliver 10 to 30 terabytes of data per night, culminating in observations of about twenty billion stars and twenty billion galaxies \citep{LSST2019}.

With the exponential increase in data acquisition, there have been complementary advances in data analysis in astronomy through the use of data mining for scientific discovery. \citet{Ball} highlighted the importance of this new era of big data astronomy and reviewed data mining efforts as an essential tool in astronomical research, without which the data from surveys would largely go unexplored, and \citet{Baron2019} provides one of the most recent summaries of modern machine learning efforts in astronomy. 

Machine learning methods can be broadly broken into two categories: supervised and unsupervised. In supervised machine learning, information is provided to the algorithm from which it must predict answers, alongside a training set of data with known answers, otherwise known as the ground truth. Whether the goal is regression for the parameters of a model or classification into known categories, this ground truth serves to teach the method how to handle the information it's given so as to return the appropriate answers. This ground truth may be split into a testing set composed of data with known answers, but which the algorithm isn't trained on, to characterize how well the algorithm performs on new data. In unsupervised learning, no such ground truth is used to train the algorithm, and instead the machine must learn exclusively from the relationships between the data.

A major focus of machine learning work in astronomy has been on efficient and accurate classification for specific science cases \citep[See for example][]{Richards2011,Bloom2012}, which has relied heavily on supervised learning. This is further exemplified by diversity of methods commonly used for classification in astronomy \citep{Ivezic2013}, the databases created and updated for surveys via machine classification \citep{Debosscher2011,Matijevic2012}, and even sponsored challenges like the Supernova Photometric Classification Challenge \citep{SN_Class_challenge} to classify supernovae based on photometry given the lack of spectroscopy for most sources, and the LSST Corporation funded Photometric LSST Astronomical Time-Series Classification Challenge \citep[PLAsTiCC;][]{PLAsTiCC} to actively direct the development of efficient classification techniques for astronomical time series data at LSST scale, while at the same time including previously unobserved variables in a catch-all class.

Classification of known phenomena cannot, however, fully explore the data space of these petabyte-scale surveys with tens of billions of targets. One of the most exciting prospects of new surveys is the potential to make discoveries of previously undetected and unknown signals. In this realm of astronomy new phenomena have, historically, been discovered serendipitously (see for example \citep{Lintott2009,Cardamone2009,Thompson2012,Wright2014}). The data volumes of modern surveys all but preclude such discoveries via happenstance. To that end, outlier detection has gained traction for the purposes of identifying potentially interesting data.

Outlier detection, alternatively referred to as anomaly detection, is the analysis technique used to identify anomalous data, data which varies from the norm beyond a threshold of expected variation.
Outlier detection is widely used outside of astronomy, a summary of outlier detection methodologies and their applications has been given by \citet{Chandola2009} encompassing different forms of anomaly detection for different purposes, like detecting network attacks \citep{Agrawal2015}, fraud \citep{Ahmed2016}, and malware \citep{Menahem2009}. At present, only individualized efforts have been applied to major astronomical surveys to identify outliers for the purpose of novelty detection with minimal overlap between studies and approach \citep[see for example, ][]{Protopapas2006,Meusinger2012,Fustes2013,Nun2016,Baron2017,Segal2019}.

This work builds on previous work in anomaly detection using the Kepler data \citep{Giles2018}, which utilized a density based clustering approach to identify outliers. Unsupervised clustering groups data based on a cluster metric (i.e. proximity or density) in feature space. Being unsupervised, no training set or classifications are trained on and the clusters are determined solely from the relationships within the data. Machine clustering has been used in astronomy to identify literal open clusters based on position, parallax, and proper motion data \citep{CastroGinard2018}, as well as to identify unique spectral classes in APOGEE data without training data \citep{GarciaDias2018}. Anomaly detection with unsupervised clustering identifies outliers as data that are unclustered.
Our previous work provided a proof-of-concept illustrating that a density based clustering approach can identify anomalies from data artifacts, rare variables, and sources of scientific interest like Boyajian's star \citep{Boyajian2016}. The lightcurve of Boyajian's star (KIC 8462852) has asymmetric dips of varying duration at seemingly random times. Its erratic behavior has been most consistent with an occulter of ordinary dust \citep{Boyajian2018}, but it was a unique signal that generated mass speculation as to its explanation and was only discovered as the data was manually scoured by citizen scientists. Our previous work using cluster-based anomaly detection identified fewer than 5 per cent of data as anomalous, dramatically focusing the search for scientifically interesting signals and other anomalies.

A natural extension of the binary determination (i.e. anomaly or not), is that of a scoring system to prioritize the study of the anomalous data. This can further facilitate discovery and enable a more efficient pipeline to identify the most egregious data artifacts. This can also be used in conjunction with known classifications to provide a likelihood that a given object is genuinely anomalous. We further that work by developing a framework to score the degree of weirdness of an object and improve computational performance. We present performance improvements, effects of parameter modifications, and outlier scores for all objects observed by the Kepler prime mission.
\section{Methods}
\label{sec:methods}
\begin{figure}
    \centering
    \includegraphics[width=\columnwidth]{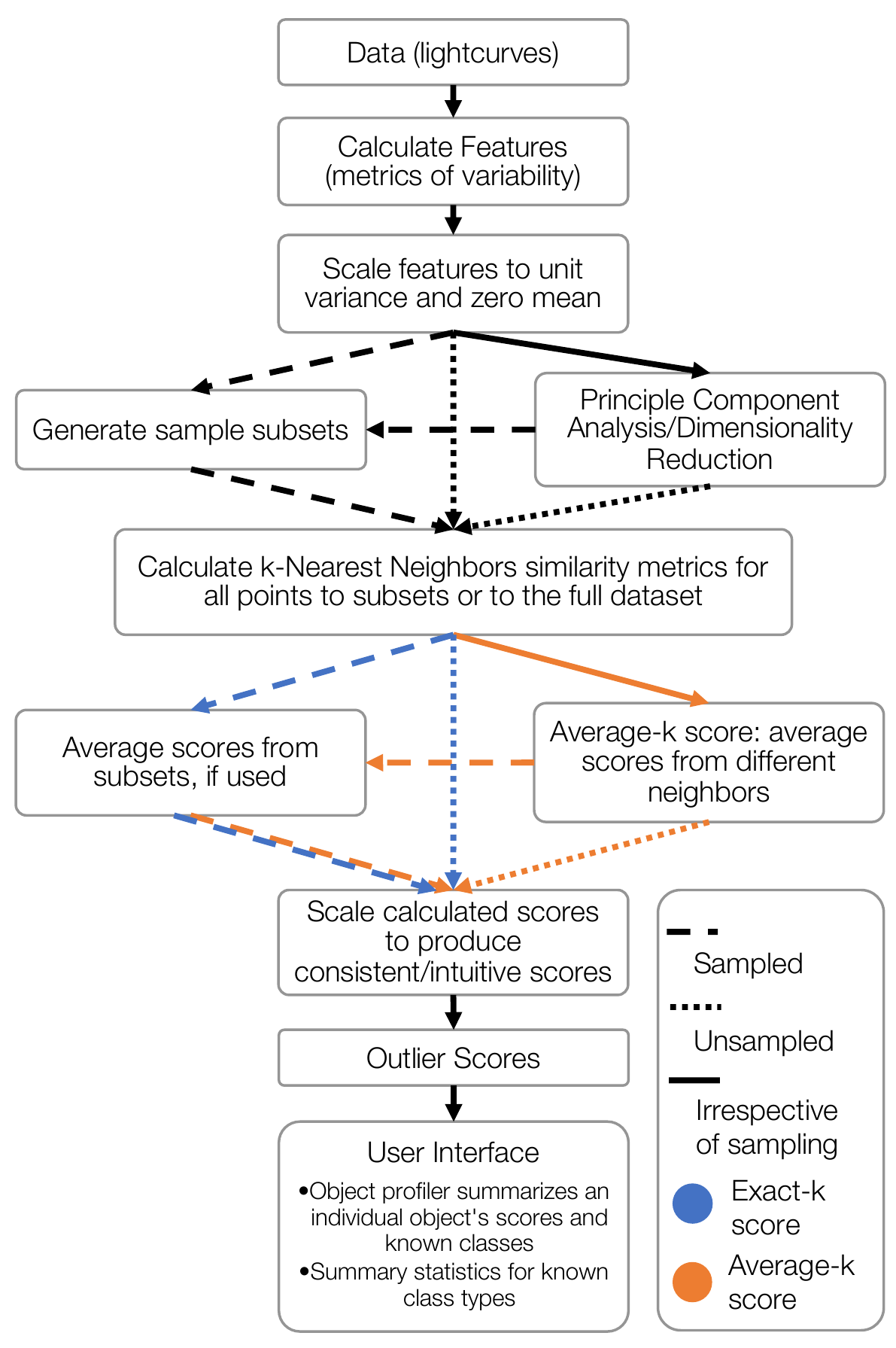}
    \caption{A visual representation of the workflow, taking lightcurves to outlier scores.}
    \label{fig:workflow}
\end{figure}

In this section we outline the methodology we have developed to produce outlier scores.
In our previous work, we assigned data one of three designations based on the clustering algorithm, DBSCAN, which creates continuous clusters based on density. In each cluster there are core cluster members which meet the density requirements of the cluster definition, edge cluster members which have core cluster neighbors but do not meet the density requirement themselves, and outliers which satisfy neither case. 
Density is based on the similarity to other data, or equivalently, the distance to specified neighbors in feature space. This may be thought of as a cluster based or proximity based definition as described in \citet{Chandola2009} and \citet{Aggarwal2013}.
In our process from lightcurve to outlier score, illustrated in Figure \ref{fig:workflow}, we first generate numerical features for each lightcurve, optionally using principal component analysis to reduce the feature space to its principal components, then calculate the distances from each point to its nearest neighbors (either each point's true nearest neighbors or to the nearest neighbors in a reference subset, see Section \ref{methods:sampling}) and finally scale the resulting distances to final outlier scores.
In Section~\ref{methods:data} we describe the data we apply our work to.
In Section~\ref{methods:features} we review the numerical features that we use to characterize each lightcurve.
In Section~\ref{methods:PCA} we detail how we use principal component analysis to reduce the dimensionality of the feature space.
In section~\ref{methods:scoring} we outline the process of determining the outlier score of each point and the different parameter spaces we explore.

\subsection{Data}
\label{methods:data}
The data we consider in this study are long-cadence photometric lightcurves from Quarters 1 to 17 of NASA's {\it Kepler} mission. We utilize Data Release 25 which reprocessed all Q0-Q17 data with the updated data pipeline \citep{KeplerArchiveManual,KeplerDR25}. We summarize here some key features of the Kepler mission and the data which are relevant to this work. and note that full specifications for Kepler can be found for instrumentation \citep{KeplerBook}
and the input catalog \citep{Batalha2010} are available via the Mikulski Archive for Space Telescopes\footnote{\footnotesize accessible online at http://archive.stsci.edu/kepler/}.

The Kepler mission was designed to observe stars in a single 105 deg$^2$ field of view (FOV) centered at R.A. = 19h22m40s and Dec = 44$^{\circ}$30'00'' from March 2009 to May 2013. Four times a year, every 3 months, the Kepler spacecraft rolled by 90 deg to re-align its solar panels, and these define epochs known as ``Quarters.'' 
With the primary goal of identifying the fraction of terrestrial exoplanets located in the habitable zone of their host star, only pre-selected stars of interest included in Kepler Input Catalog (KIC) had continuous data downloaded in the Kepler prime mission \citep{Batalha2010}.
The Kepler Mission achieved about 30ppm for solar type stars \citep{Gilliland2011,Gilliland2015}. Stars where an exoplanetary transit signature of around 100ppm is impossible to detect (i.e., giants, stars fainter than 16th mag, stars in overcrowded fields) were omitted from the target list. Of the roughly half-million targets in the FOV brighter than 16th magnitude, only ~30\% were targeted. Beyond the primary target list, additional high priority targets included all known eclipsing binaries in the FOV ($>$600), all members of open clusters in the FOV, and the nearest main sequence stars.
For this work we have utilized the full KIC long-cadence observations which are composed of 270, 6.02s exposures per observation (about half an hour per observation), and over 4,000 observations per epoch. 

The calibration pipeline for Kepler lightcurves, like target selection, is optimized toward the goal of identifying exoplanetary transits. The pipeline does not attempt to remove all artifacts from lightcurves. The primary means to identify and clean instrumental signatures and systematic errors, the Presearch Data Conditioning pipeline, corrects or removes affected data where possible, but is known to perform poorly for systematics that are non-temporally correlated between targets.
Details on the Kepler data processing are available in the Kepler Data Processing Handbook \citep{KeplerDataProc}, and known ongoing phenomena are documented in the Kepler Data Characteristics Handbook \citep{KeplerDataChar}. We do not attempt to remove any remaining artifacts prior to our analysis as the identification of artifacts as anomalies is one application of this work.

\subsection{Features}
\label{methods:features}
We calculate 60 numerical features from the lightcurve data describing the behaviour of the flux over time. Our feature set is largely derived from \citet{Richards2011}, who demonstrated the utility of numerical features for classifying variable stars with lightcurves of differing sparsity and noise. While the data we use from Kepler is consistently well sampled and precise, we include the step of feature generation to make this approach more generally applicable. We also use additional features developed specifically for use with the Kepler data \citep{Walkowicz2014} which have a focus on efficient calculation for large sets of data. 
All features are numerical and treated as continuous, though some features are inherently discrete. Each feature is offset to have a mean of zero and scaled to unit variance prior to principal component analysis and scoring. We summarize the features themselves, the motivation to use features, and their utility in greater detail in our previous work \citep{Giles2018}.

We note here that we have made a small number of corrections and modifications to the calculation of most features from the prior work, documented in more detail in the Outlier Scoring branch of this work's GitHub repository
\footnote{\footnotesize{ 
https://www.github.com/d-giles/KeplerML}}, and have made significant improvements to the computation of the features heavily utilizing Anaconda's Numba package \citep{Lam2015}. 

\subsection{Feature Preprocessing}
\label{methods:fts_math}
Here, we describe how features are preprocessed for scoring after initial calculation and formally define some terminology. We use define the following terms:
\begin{description}
\item `feature' to refer to the concept of a particular measurable or calculable value from a lightcurve,
\item `calculated features' as individual calculated, or measured, values, $m$, 
\item `feature vector', the full set of calculated values for a particular object,  eq. \ref{eq:featureVector},
\item `feature set' as the full set of values for a particular feature, $f_\phi$, Equation \ref{eq:featureSet}.
\end{description}
We calculate scores by quarter, $\mathcal{Q}$, the set of all objects for a single quarter of Kepler data. 
First, we calculate each feature, $m$, for each element, $i$ of $\mathcal{Q}$ and store these values together as a feature vector, $\mathbfit{F}_i$.
\begin{equation}\label{eq:featureVector}
    \mathbfit{F}_i = 
    \begin{bmatrix}
        m_{i,1} & \dots & m_{i,D}
    \end{bmatrix},\\
\end{equation}
On calculating features for all elements in $\mathcal{Q}$, we store all feature vectors as matrix, $\mathbfss{F}$, the set of all feature vectors, or equivalently, the set of all calculated feature values.
\begin{equation}\label{eq:featureMatrix}
    \mathbfss{F} =
        \begin{bmatrix}
        \mathbfit{F}_1 \\
        \vdots \\
        \mathbfit{F}_N
     \end{bmatrix}=
    \begin{bmatrix}
    m_{1,1} & \dots &m_{1,M} \\
    \vdots & \ddots & \vdots \\
    m_{N,1} & \dots &m_{N,M}
    \end{bmatrix},
\end{equation}
where $N$ is the number of elements in $\mathcal{Q}$ (the number of lightcurves in the quarter), and $M$ is the number of features calculated for each lightcurve.

The data are scaled to unit variance and zero mean by feature, $\phi$, per Equation~\ref{eq:scaledFeatureVector}.
\begin{equation}\label{eq:scaledFeatureVector}
    \mathbfit{X}_i = \{\frac{m_{i,\phi}-\mu_\phi}{\sigma_\phi}|m_{i,\phi}\in \mathbfit{F}_i\},
\end{equation}
where $\sigma_\phi$ is the standard deviation, and $\mu_\phi$ the mean, of each feature set, $\mathbfit{f}_{\phi}$ (Equation \ref{eq:featureSet}).
A feature set is the full set of calculated features for a given feature.
\begin{equation}\label{eq:featureSet}
    \mathbfit{f}_{\phi} =
    \begin{bmatrix}
        m_{1,\phi}\\
        \vdots\\
        m_{N,\phi}
    \end{bmatrix}.\\
\end{equation}

In the case of using the full feature set, this includes all 60 calculated features and in principal component reductions, as described in Section~\ref{methods:PCA}, the feature set is composed of the principal components.

The preceding process is performed using ScikitLearn's preprocessing module and features are stored as full feature matrices (eq. \ref{eq:featureMatrix}) as pickled Pandas Dataframes.

\subsection{Scoring}
\label{methods:scoring}
In our previous work, we demonstrated the utility of a density based outlier detection approach on Kepler data. We expand on that work by defining an outlier score to better describe identified outliers. A score can provide better guidance when seeking to identify the most interesting or egregious data anomalies and forms a basis for prioritizing followup investigation.
When used with existing knowledge of nominal data and known variability classes, anomaly score distributions can be used to provide a probability estimate that a given object is uniquely anomalous versus belonging to a particular class \citep{Gao2006}. 
We perform scoring by quarter and scores are relative to the other objects within a quarter, not necessarily to objects outside of that quarter. In Section~\ref{methods:reference}, we suggest a way to scale the scores so that they may be comparable across quarters.

\subsubsection{General Score Definition}
\label{scoring:maths}
For each quarter we produce a set of scores, $\mathcal{S}$, for all members of the quarter. 
The outlier score we calculate for a given point, is a measure of the sparsity of neighboring points around the point of interest. We determine the sparsity by calculating a similarity metric to a point's closest neighbors, the distances in the feature space where each point's coordinates is given by its feature vector \citep{Anguilli2002,Chandola2009,Ram2009,Upadhyaya2012,Aggarwal2013}.

We define the k-distance for each point in a quarter, $\mathcal{Q}$, as the distance to the k-th nearest neighbor in a reference set, $\mathcal{Q'}$, where $\mathcal{Q'} \subseteq \mathcal{Q}$. The matrix, $\mathbfss{D}$, contains the sets of distances for all objects in $\mathcal{Q}$ to their nearest $n$ neighbors in $\mathcal{Q'}$.
\begin{equation}\label{eq:distanceMatrix}
    \mathbfss{D}=
    \begin{bmatrix}
        d_{1,1} & \dots &d_{0,n} \\
        \vdots & \ddots & \vdots \\
        d_{N,1} & \dots &d_{N,n}
    \end{bmatrix},
\end{equation}
where $d_{i,j}$ is the distance to the $j^{th}$ nearest neighbor of the $i^{th}$ element.

As part of this work we test the effect of variations on the scoring definition, including different values of k and averaging over several values of k. For every $k$ we are producing a unique score for each object, and larger k correspond to further neighbors and greater distances. Different choices of k, therefore, provide scores on different scales. Averaging over several values of $k$ is a basic ensemble of different scores. \citet{Zimek2014} and \citet{Aggarwal2017Ensembles} discuss challenges in outlier ensembles and emphasize that some form of normalization needs to be applied to ensure that scores to be combined are on the same scale. To normalize the different scores we MinMax scale each column of $\mathbfss{D}$ from zero to one.
These scaled distances are exact-k scores, and form the building blocks for averaged scores.
In the case that they are used as building blocks, we refer to these scaled k-distances as k-protoscores, $p_{i,k}$. 
\begin{equation}\label{eq:protoscore1}
    p_{i,k}=\frac{d_{i,k}-d_{\text{min},k}}{d_{\text{max},k}}.
\end{equation}
For exact-k scores $k_0=k=K$ and Equation~\ref{eq:protoscore1} is equivalent to the k-protoscore. For an average score, $k_0\leq k\leq K$, we take the mean of k-protoscores to produce the k-averaged protoscore, $\bar{p_i}$. Equation~\ref{eq:protoscore2} is the general form of the protoscore.
\begin{equation}\label{eq:protoscore2}
    \bar{p_i} = \frac{1}{K-k_0+1}\sum_{k=k_0}^{K} p_{i,k}.\\
\end{equation}
For protoscores computed with respect to randomly sampled subsets we mean average the protoscores for each reference sample for the final protoscore. These protoscores are then linearly scaled into final scores as described in Section~\ref{methods:reference}.

As regards the values of k explored, a heuristic for DBSCAN clustering was suggested by \citet{Ester1996} (the original DBSCAN paper) to choose k as 4 and determine epsilon based on where the graphical elbow occurs for distance to that 4th neighbor. \citet{Ester1996} suggests that there is little gain or difference beyond k=4 in terms of clustering performance, however, our previous work \citep{Giles2018} found k=4 to be insufficient to produce an epsilon of sufficient size for clustering.
We use k=4 as a starting point for scoring and increased from there. 
We considered 10 exact scores from $4\leq k<14$. 
For $k\geq14$ we consider k-average scores exclusively to better understand the effects of the scale of k, and to downplay variance between individual values for k. We consider averages for sets of 10 from $14\leq k<104$, and averages for sets of 100 from $4\leq k<1004$.

\subsubsection{Variations on Score Definition}
\label{scoring:parameters}
For Quarter 1 of the Kepler data, we test the sensitivity of scores with respect to varying k and averaging over different k ranges (detailed in Section \ref{scoring:maths}), scoring with respect to subsets versus the entire quarter (see Section \ref{methods:sampling}), removing potentially problematic features (see Section \ref{methods:problems}), and using PCA to reduce dimensionality (see Section \ref{methods:PCA}). We produce scores for all variations on a single quarter, examining the effects of changing each parameter independently by assessing the change in scores for a given change in parameters.

Given the unsupervised nature of anomaly detection, we do not have a ground truth to evaluate the performance and to optimize the parameters for scoring. Further, \citet{Aggarwal2013} emphasize the risk of evaluating outlier scoring performance with any internal measurement (i.e. comparing to another metric for the local density for each point), as any choice of evaluation can easily be biased towards the particular choice of algorithm, and suggests an imperfect external validation is to use known rare classes as a ground truth for anomalies. 
Treating rare object types (<1000 members in the KIC as given by the SIMBAD database) as the ground truth, we evaluate the ability of our methodology to identify the most rare classes of objects by examining the area under the receiver operating characteristic (ROC) curve using the score as a threshold. The ROC curve plots the true positive rate versus the false positive rate. For a random classifier, the curve would follow the diagonal and the area under the curve (AUC) would be 0.5, while a perfect predictor would have an AUC of 1.0.
In this case, we take true positives to be rare variables identified as outliers, and false positives as more common object types.

\subsection{PCA Reduction}
\label{methods:PCA}
In addition to scoring the full feature set, we have implemented principal component analysis (PCA) for dimensionality reduction \citep{Jolliffe2002} prior to outlier scoring. PCA is used to identify orthogonal axes to maximally explain variance, and it has been used widely in astronomical analysis \citep[provides several examples, though many more exist]{Baron2019}. The use of PCA dimensionality reduction has two primary motivations in this work. The first motivation is to address the effects of the curse of dimensionality. The high dimensionality affects outlier detection in a few key ways; in higher dimensional data, points become more evenly disbursed directly, potentially affecting our distance-based scoring metric.
Second, the features we use, described in Section \ref{methods:features}, have been shown to be useful in distinguishing variable classes, but are also heavily correlated in some cases. In the distance based scoring approach we use, this can artificially emphasize extrema in correlated features. \citet{Aggarwal2013} suggests that using PCA to reduce correlation between dimensions may improve distance based scores in large distributions.
Additionally, higher dimensional data require additional computation time and resources proportional to the number of dimensions being considered. Though, as the number of dimensions is typically orders of magnitude smaller than the number of objects being considered, as is the case in our work, scalability is dominated by the number of objects.
To generate reductions for a single quarter, we first scaled each feature set to zero mean and unit variance using StandardScaler, then used the PCA module from Scikit-learn \citep{scikit-learn} to create reductions which explained 90, 95, and 99 per cent of the variance (PCA90, PCA95, and PCA99 respectively).

\subsection{Sampling}
\label{methods:sampling}
Beyond the advancements we have made in computational performance from improved feature processing and data handling, we are interested in using sampling techniques to further enable scalability of these methods. Calculating distances between all points scales as $O(N^2)$, where N is the number of datapoints in the set, though the use of Scikit Learn's NearestNeighbors module allows this to scale closer to $O(N\log{N}$ by pruning the data and calculating only a subset of the distances. Calculating distances to a reference subset of data can dramatically improve on this, scaling as $O(sN)$, where s is the size of the subset. 

When sampling, we draw multiple random subsets, $\mathcal{Q'}$, from a quarter, $\mathcal{Q}$ where $\mathcal{Q'}\subset\mathcal{Q}$. A protoscore for each element of Q is calculated based on the nearest neighbors in each subset $\mathcal{Q'}$ per Equation \ref{eq:protoscore2}, then protoscores for each subset are mean averaged for the final protoscores. In this work, we evaluate the scores produced by an average of 10 subsets of 10k points, 50 subsets of 1k points, and 10 subsets of 1k points.
\subsection{Problematic Features}
\label{methods:problems}
We note that for two features, the flatness and roundness ratios of maxima to minima, a divide-by-zero error can occur if there are no minima in a lightcurve. Rather than remove lightcurves where the error would occur, we assign a large dummy value rather than removing the data since it was unknown whether it would be a common or uncommon issue. If common the data may cluster, and if uncommon the data would appear as outliers. We examine the effects of scoring without these two features in Section \ref{results:problem_fts}.
\subsection{Score Scaling}
\label{methods:reference}
The scores produced by this work are fundamentally similarity metrics, or rather dissimilarity metrics, representing the relative oddity of each object against other objects in the same dataset.
In Section \ref{scoring:maths} we discussed the importance of score normalization for comparing scores from different methods. Propagating the normalization scheme we used for calculating k-average scores, we use a zero to one scale for our primary scoring by quarter to indicate the degree of outlying nature (one being most outlying) of each object. We MinMax scale the final set of protoscores calculated by Equation \ref{eq:protoscore2}.

However, each quarter has, and all datasets more broadly have, a different greatest outlier which is assigned the most outlying score of one. This can be misleading when comparing scores across quarters or datasets. In the interest of comparing scores across quarters, and potentially across similar datasets and surveys, we propose scaling to a reference as a viable method for comparison across similarly populated datasets. 
We cannot know the most exotic forms outliers will take in a given dataset, but we do know the staggering majority of targets in the Kepler data primarily present small scale variability as observed locally, brightening and dimming within 1\% of the median flux. As such, we use a reference for scaling that is known to be relatively odd compared to the bulk of the data. We use a sinusoid to produce reference features. The reference data is produced to mimic a lightcurve for a single quarter of the Kepler data, simulating a 13 week duration with a data point every 30 minutes. The sinusoid is given a one week period and 50\% amplitude flux deviation. We calculated features for the simulated lightcurve and appended the reference into all quarters of data with the ID 'sinusoid'.
Following scoring per Section~\ref{scoring:maths}, we linearly scale the scores such that the sinusoid reference has a score of one.
Including the reference feature vector as though it were actual data when calculating scores carries a potential to affect the scores of physical data. It could be the $k^{th}$ neighbor of actual data, thereby generating a score based on artificial data. However, as a singular point this only affects exact-k scores and must exist between the $k\pm1$ genuine neighbors, the risk is considered minimal. 

\section{Results}
\label{sec:results}
Here, we present comparisons between different score definitions and results from scoring the Kepler light curves.
First, without a known ground truth or another external validation method available, we have sought to qualitatively investigate distance-based scoring. We summarize the results of the different scoring methods as applied to Quarter 1 of the Kepler data in Section~\ref{results:scoring-comp}, comparing the differences between scoring methodologies as detailed in Section~\ref{methods:scoring}. 
Next, we present a summary of the scoring results for all quarters of the Kepler data as well as a sample of the most outlying lightcurves in Section~\ref{results:scoring}.
Finally, we discuss placing outliers in context including the results of scaling scores with respect to a common reference between different quarters in Section \ref{results:context}

As we discuss scores produced from a particular set of parameters we use the following conventions. First, we indicate whether we are scoring on the feature data or on PCA components, next we specify the sampling used, if any, and finally specify the which neighbor or range of neighbors are used for scoring. The sampling is given either as `full', where no sampling is used, or as number of subsets\x~size of subsets. The neighbor to which distance is calculated for exact-k scores is referred to as k=$value$, and for k-average scores the range of neighbors is referenced as \krange{min value}{max value}.
\subsection{Scoring Comparisons}
\label{results:scoring-comp}
Here we take a detailed look at the scoring results for the differing score methodologies, comparing the resultant scores for Quarter 1 of the Kepler data. 
\begin{figure}
	\includegraphics[width=\columnwidth]{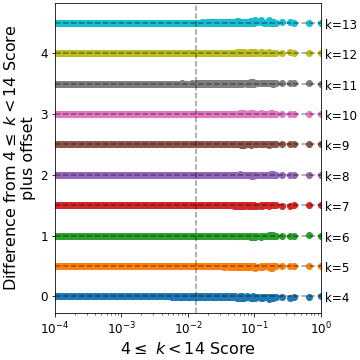}
    \caption{The difference between exact-k scores for k=4 to k=13 are plotted here against the k-average score \krange{4}{14} for the full feature set of Quarter 1 data. There are only minor variations between any given exact-k score and the average-k score. Differences are offset from one another and horizontal lines are added at increments of 0.5, the dashed vertical line separates the top 1000 outlier scores from the \krange{4}{14} to the right, and the remaining 155,126 scores to the left.}
    \label{fig:ex-4-13}
\end{figure}
\begin{figure}
	\includegraphics[width=\columnwidth]{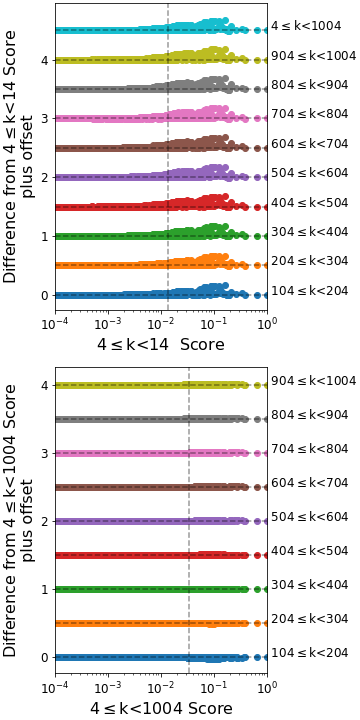}
    \caption{Here we examine the effect of higher choices of k. On the top, Scoring is based on the average of different sets of 10 neighbors from k=4 to 103 with the topmost score the average over the whole 100 neighbors. On the bottom, scoring is based on the average of different sets of 100 neighbors from k=4 to 1003 Scores calculated against the full set of data. The lowest valued sets of neighbors tend to have lower intermediate scores, but are consistent on the end points. There is little to no variation in scores between higher sets of neighbors. The vertical line in both plots indicates a separation of the top 1,000 most outlying points to the right.}
    \label{fig:highk}
\end{figure}
\subsubsection{Varying k}
First, we consider how choice of k impacts scoring. In Figure~\ref{fig:ex-4-13} we look at small values of k scored using the full feature set against the full quarter. 
In this figure, we plot the difference of each set of exact-k scores (k=4 to k=13) from the \krange{4}{14} k-average score against the \krange{4}{14} k-average score. We see almost no change from one k to another for these small values of k. The dashed vertical line in the figure demarcates the top 1,000 outlier scores to the right. Figure~\ref{fig:highk}, however, illustrates that scoring to much greater neighbors does have more of an impact on scores from low neighbors to high neighbors, but to be relatively invariant at high neighbors. This figure shows the scores averaged from \krange{4}{1004} in ranges of 100. On the left, the difference from the \krange{4}{14} scores are shown, and on the right the difference from the \krange{4}{1004} scores. 

\subsubsection{Scoring to Subset Samples}
\label{results:sampling}
\begin{figure}
	\includegraphics[width=\columnwidth]{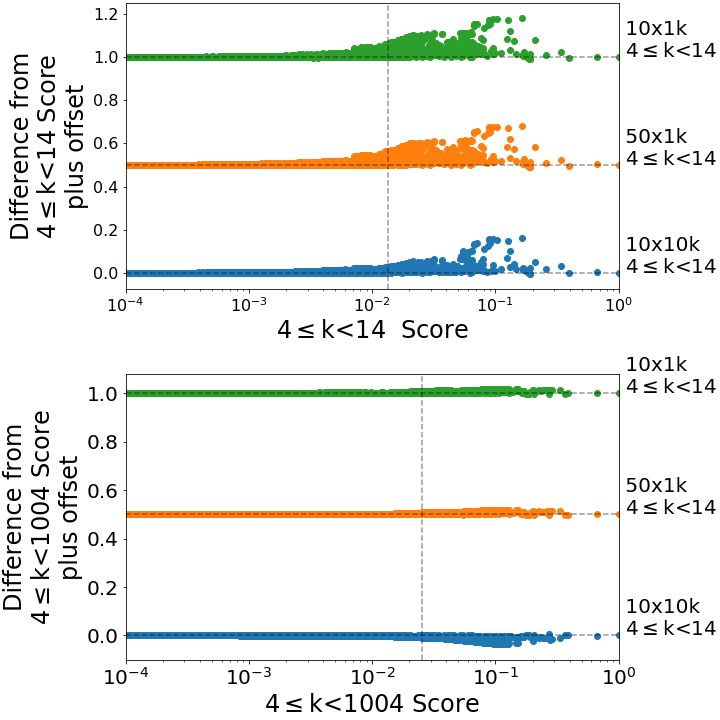}
    \caption{In the top figure, we plot the difference between \krange{4}{14} scores as scored to different samples from the \krange{4}{14} scored to the full quarter. In the bottom figure  we plot the difference between \krange{4}{14} k-average scores as scored to different samples from the \krange{4}{1004} k-average scores as scored to the full quarter. The \krange{4}{14} scores from the sampled reference strongly resemble the high-k scores.}
    \label{fig:sampling-params}
\end{figure}
In Figure~\ref{fig:sampling-params} we look at the effect of sampling on outlier scores. In the top plot, we show the difference for the \krange{4}{14} scores for different sampling. In the bottom plot we show the difference between the \krange{4}{14} scores from sampling against the \krange{4}{1004} scores and see that sampling effectively mimics high-k scores. The specific choices for sampling impact the scoring results minimally, which appears to be consistent with the score invariance between high-k scores we noted earlier.
Sampled scores strongly resemble large value neighbor scores and are computed in a fraction of the time. Using the NearestNeighbors module of SciKit Learn, with the `ball-tree' algorithm and the built in parallel processing score calculation on 48 2.70 GHz Intel Xeon CPUs of exact-k scores took 10-15 minutes to score in reference to the full dataset, 10-12 minutes to score to 10 subsets of 10k points, 5-7 minutes to score to 50 subsets of 1k points, and \textasciitilde 1 minute to score to 10 subsets of 1k points.
\subsubsection{Scoring on Different Features}
\label{results:PCA}
\begin{figure}
	\includegraphics[width=\columnwidth]{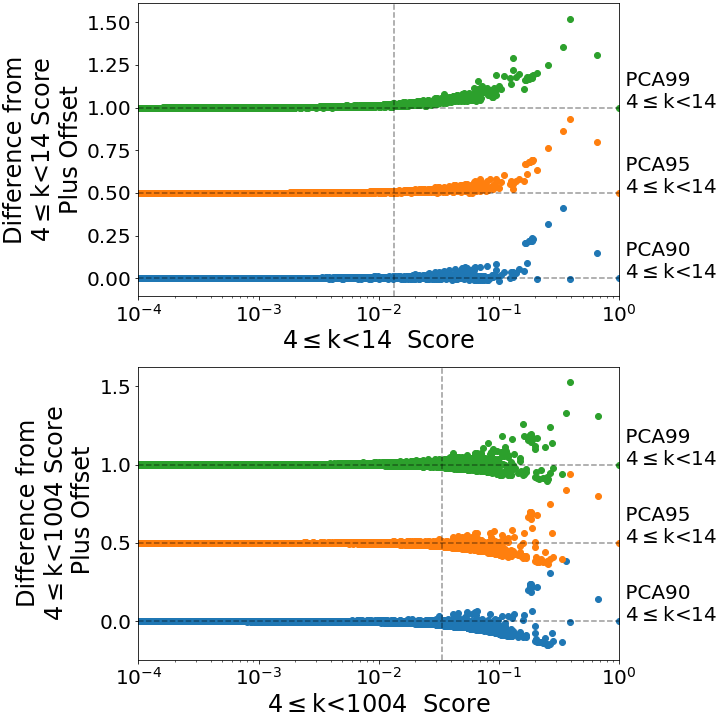}
    \caption{Score differences of each \krange{4}{14} score from the PCA reduction with respect to the \krange{4}{14} score from the full feature set are plotted against the \krange{4}{14} scores from full feature set in the top plot, and against the \krange{4}{1004} scores in the bottom plot. Scoring based on the PCA reductions has a pronounced effect on scores.}
    \label{fig:PCA}
\end{figure}
In Figure~\ref{fig:PCA} we can see the effect of using PCA reductions for scoring is much more pronounced than the choice of neighbor or sampling. In the top plot we compare the \krange{4}{14} scores from the PCA reductions to the low k \krange{4}{14} scores from the full feature set and to high k \krange{4}{1004} scores in the bottom plot. The scores based on PCA components do not resemble either high or low values of k, impacting the scores distinctly, particularly the most outlying points as most variation appears to the right of the vertical line which demarcates the top 1,000 most outlying scores per the x-axis.

\begin{figure}
    \centering
    \includegraphics[width=\columnwidth]{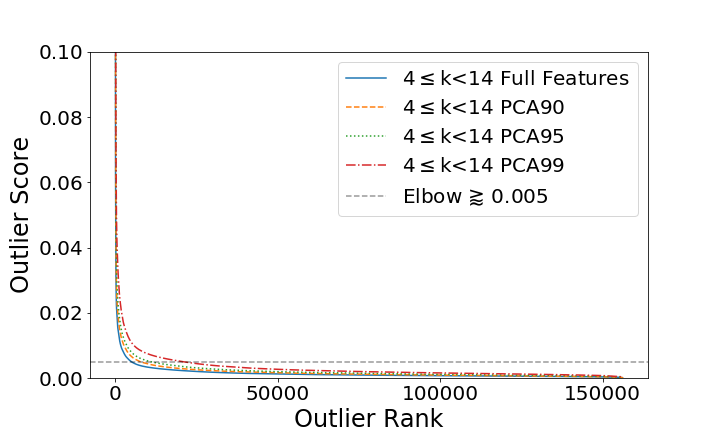}
    \caption{The elbow in the score versus rank plot acts as a heuristic measure of the cutoff score between normal points and outlying points. We plot the outlier score versus the corresponding rank for the \krange{4}{14} scores from the full feature set and the PCA reductions, where the most points are ranked from most to least outlying, i.e. the most outlying point has a rank of zero. For the \krange{4}{14} scores from the full feature set, we see the elbow at around 0.005 with around 5k outliers above and to the left of the elbow. The elbows for the PCA reductions are comparable.}
    \label{fig:score_elbow}
\end{figure}

\begin{figure}
    \centering
    \includegraphics[width=\columnwidth]{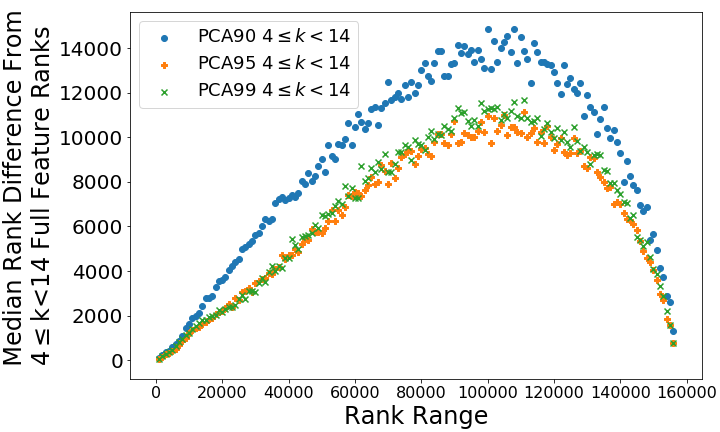}
    \caption{The median absolute differences between outlier ranks of \krange{4}{14} scores from the PCA reductions to those calculated from the full feature set. Low ranks correspond to more outlying points. Points are sorted by scores of the full feature set and ranges are sets of 1,000 points, each x-component corresponds to the uppermost rank of each range. The ranks converge for the most and least outlying ranges indicating that the most outlying points are consistently identified as such. The intermediate scores appear to vary in rank significantly with minor score changes as the scores for intermediate points are so similar.}
    \label{fig:PCA_rank_diff}
\end{figure}

As outlying lightcurves are the data in which we're most interested, we examine the extent to which these score variations affect the relative ranking of outlying points.
As a heuristic measure of the cutoff between normal and outlying data, we identify the location of the elbow in the score versus rank plot, as per the suggested heuristic for finding the cutoff distance for DBSCAN \citep{Ester1996}. In Figure \ref{fig:score_elbow}, we see an elbow around a score of 0.005 for the \krange{4}{14} full feature scores, and comparable elbows for the PCA reductions. This elbow separates the roughly 5,000 most outlying points from the remaining points. In Figure \ref{fig:PCA_rank_diff}, we show that the relative ranks of the most outlying points, those 5,000 past the elbow in Figure \ref{fig:score_elbow}, are consistent despite the scoring differences between PCA reductions. In the intermediate region between the most and least outlying points, there is more variance as the ranked order of a given lightcurve can change dramatically from small score changes where the scores are very similar. It also appears that the least outlying points are consistently identified as such at the highest outlier ranks. However, we are primarily concerned with the relative ranking of outliers which are more or less invariant to the reduction, even despite the greater score variation we can see in Figure \ref{fig:PCA}.

\subsubsection{Problematic Features}
\label{results:problem_fts}
\begin{figure}
    \centering
    \includegraphics[width=\columnwidth]{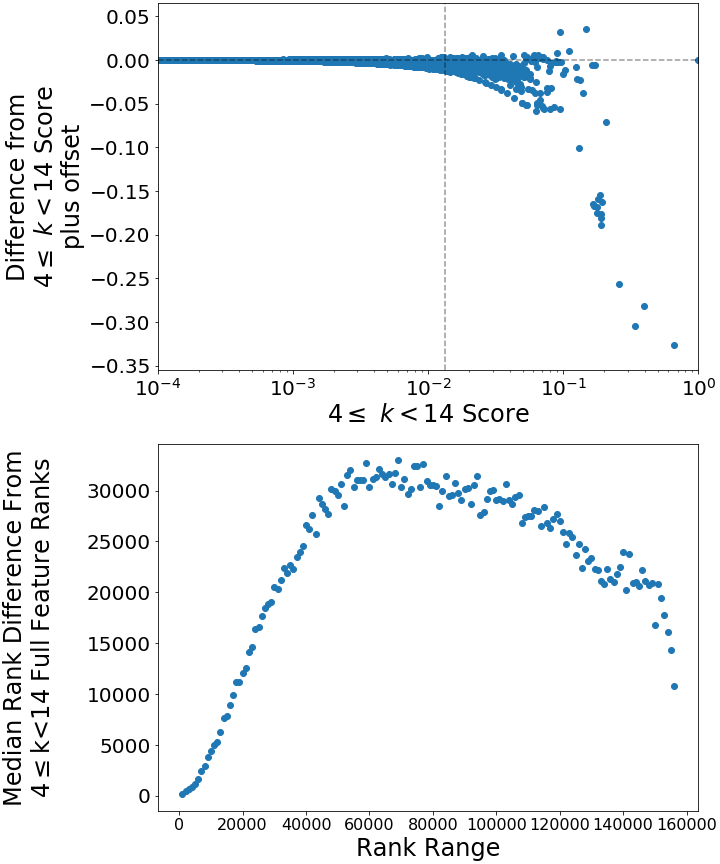}
    \caption{Scoring on the features minus the ratios of the flatness and roundness of maxima to minima has a marked effect on scores. The most outlying points, however, still have good consistency in relative ranking compared to the rest of the points.}
    \label{fig:ft_sub}
\end{figure}

We provide some of the most outlying lightcurves identified in this work in Appendix \ref{appendix:lightcurves}, and in Figure \ref{fig:Q1topten} we see that three of the top ten outliers in Quarter 1 of the \krange{4}{14} k-average scores for the full feature set are long period variables. We also see that each has no minima which affects two features in particular, the flatness and roundness ratios of maxima to minima.
Scoring without the flatness and roundness ratios has a dramatic effect on the scores, largely decreasing outlier scores for select objects as can be seen in the top plot of Figure \ref{fig:ft_sub} and in the different top outliers in Figure \ref{fig:Q1topten_ftsub}. However, similar to the PCA reductions, on average the most outlying points tend to remain so, as can be seen in the bottom plot of Figure \ref{fig:ft_sub}.

\begin{figure}
    \centering
    \includegraphics[width=\columnwidth]{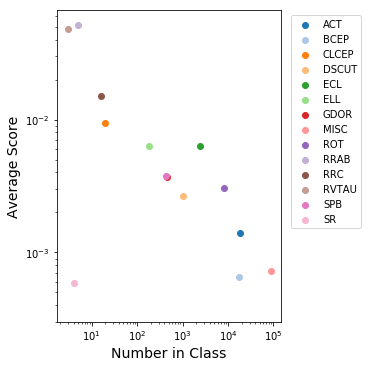}
    \caption{Q1 average score by Debosscher class, plotted against class count. See Table \ref{tab:debosscher_classes} for specific counts and scores.}
    \label{fig:scorevscount}
\end{figure}

\subsection{Outlier Scores}
\label{results:scoring}
\begin{figure}
    \centering
    \includegraphics[width=\columnwidth]{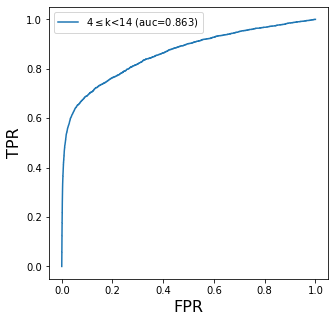}
    \caption{The receiver operating characteristic plots the true positive rate against the false positive rate to evaluate the performance of a classifier. In this case, we treat relatively rare object types as true positive outliers and evaluate the effectiveness of the scores to identify the most rare objects.}
    \label{fig:roc}
\end{figure}
We have produced scores for all 2.8 million individual light curves for \textasciitilde 200k individual sources for hundreds of variations on scoring parameters, dimensionality reductions, and sampling choices. Given the results presented in Section \ref{results:scoring-comp} that scores and relative ranks for the most outlying points are comparable within a quarter, we present here only the \krange{4}{14} scores for the full feature set without sampling. These scores are scaled from zero to one where the most outlying object in the quarter has a score of one. This scale allows for a rudimentary "at-a-glance" understanding of the relative outlying magnitude of an object with respect to the rest of a quarter where the closer to one a score is, the more outlying that object is. Though, as discussed in Section \ref{results:PCA}, the cutoff between outlying and non-outlying can be quite small ($\leq 0.005$ for \krange{4}{14} k-average scores).
In each quarter, the most outlying lightcurve has extreme behaviors stemming from a mix of astrophysical processes as in SU Uma Cataclysmic Variables (KICs 7446357 and 8415928) and Eruptive Variables (KIC 7199037), and data artifacts resulting in negative flux values (KICs 5944707, 7199774, 9594451,8451871,10057002)
or apparent recalibration mid-observation (KICs 5262664, 11199438). We plot the most outlying object from each quarter in Appendix \ref{appendix:lightcurves}, alongside the top ten most outlying objects of Quarter 1 in Figure \ref{fig:Q1topten}.

With sampling methods and appropriate features, anomaly detection and scoring with kNN reliably identifies the same data as the most anomalous within a large set. When compared to databases of known variability types, including eclipsing binaries and heartbeat stars \citep{Kirk2016}, Kepler Objects of Interest \citep{Coughlin2016}, solar flares \citep{Davenport2016}, stellar variability \citep{Debosscher2011} and the Simbad database \citep{Wenger2000}, we have found that every variable class skews towards higher scores, whereas the unclassed stars (the majority of objects observed by the Kepler prime mission) skew towards lower scores. 
See, for example, Table \ref{tab:debosscher_classes} and Figure \ref{fig:scorevscount} which illustrate that the average score decreases with class population for classes identified by \citet{Debosscher2011}. The 4 semi-regular variables identified by \citet{Debosscher2011} with Quarter 1 scores are a notable outlier to this trend, and we note that their variability is constrained to $\lesssim 0.1\%$ of the median flux. In the absence of ground truth for anomaly detection, this lends strong support to the effectiveness this anomaly detection method to highlight objects of greater interest. 

As described in Section \ref{scoring:parameters},
treating rare object types as the ground truth we evaluate methodology's ability to identify the most rare classes of objects by examining the area under the ROC curve. In Figure \ref{fig:roc} we show the ROC curve for the \krange{4}{14} k-average scores for the full feature set using the score  as a threshold. 
While we only show the ROC curve for the \krange{4}{14} scores, we note that the curve is nearly identical for all other score variants. 
\begin{table*} 
	\centering
	\caption{Outlier scores min-max scaled from zero to one. These scores are average-k scores, \krange{4}{14}.}
	\label{tab:minmaxscores}

	\begin{tabular}{lrrrrrrrrrrrrrrrrr}
        \hline
         & Q1 & Q2 & Q3 & Q4 & Q5 & Q6 & Q7 & Q8 & Q9 & Q10 & Q11 & Q12 & Q13 & Q14 & Q15 & Q16 & Q17 \\
        KIC & $10^{-3}$ & $10^{-3}$ & $10^{-3}$ & $10^{-3}$ & $10^{-3}$ & $10^{-3}$ & $10^{-3}$ & $10^{-3}$ &         $10^{-3}$ & $10^{-3}$ & $10^{-3}$ & $10^{-3}$ & $10^{-3}$ & $10^{-3}$ & $10^{-3}$ & $10^{-3}$ &     $10^{-3}$ \\
        \hline
        757076 & 1.23 & 1.13 & 1.06 & 0.38 & 0.76 & 0.52 & 0.38 & 0.19 & 0.89 & 1.34 & 1.11 & 0.66 & 2.47 & 3.31     &     1.09 & 0.80 & 0.53 \\
        757099 & 5.94 & 10.47 & 5.47 & 4.72 & 5.69 & 3.35 & 2.04 & 2.35 & 3.41 & 4.72 & 3.83 & 6.54 & 7.58 &     9.02 &     7.06 & 4.48 & 2.58 \\
        757137 & 1.09 & 1.11 & 0.60 & 1.13 & 1.05 & 0.61 & 0.65 & 0.64 & 0.64 & 0.92 & 1.18 & 1.03 & 1.21 & 1.07     &     0.81 & 1.27 & 0.76 \\
        757280 & 1.66 & 1.08 & 1.59 & 0.66 & 1.05 & 0.73 & 0.66 & 0.64 & 1.30 & 2.67 & 0.56 & 0.68 & 0.93 & 0.78     &     0.67 & 1.15 & 0.39 \\
        757450 & 6.02 & 3.65 & 4.03 & 2.36 & 5.57 & 2.56 & 2.75 & 2.23 & 4.18 & 3.88 & 2.67 & 3.74 & 6.56 & 5.11     &     3.39 & 1.89 & 1.73 \\
        ... & ... & ... & ... & ... & ... & ... & ... & ... & ... & ... & ... & ... & ... & ... & ... & ... &     ... \\
	    \hline
	\end{tabular}
Full, machine readable table available at https://dx.doi.org/10.7910/DVN/H5QXUL.
\end{table*}

\begin{table}
    \centering
    \caption{Quarter 1 average scores for Kepler objects classified by \citet{Debosscher2011}.}
    \label{tab:debosscher_classes}
    \begin{tabular}{lrr}
        \hline
        Class & Count & Average Score \\
        \hline
        ACT & 20,862 & 0.00174 \\
        BCEP & 19,187 & 0.00090 \\
        CLCEP & 21 & 0.01159 \\
        DSCUT & 1,060 & 0.00325 \\
        ECL & 2,544 & 0.00772 \\
        ELL & 200 & 0.00704 \\
        GDOR & 488 & 0.00429 \\
        MISC & 96,745 & 0.00100 \\
        ROT & 8,654 & 0.00395 \\
        RRAB & 6 & 0.06685 \\
        RRC & 16 & 0.01893 \\
        RVTAU & 3 & 0.06561 \\
        SPB & 466 & 0.00433 \\
        SR & 4 & 0.00074 \\
        \hline
    \end{tabular}
\end{table}

\subsection{Contextualizing Anomalies}
\label{results:context} 
We have used a convention of scaling scores by quarter from zero to one to allow a quick determination of how outlying a point is in a given quarter. However, the same lightcurve could have very different scores if compared to different quarters as the most outlying object, and its distance to its neighbors in feature space, changes from quarter to quarter, the different most outlying lightcurves for each quarter can be seen in the figures in Appendix \ref{appendix:lightcurves}.
To supplement this, we have scaled each quarter with respect to an artificial signal as described in Section \ref{methods:scoring}. We scale the scores such that the reference lightcurve has a score of one, so there is no explicit maximum value for the most outlying data. With a common reference and assuming a similar population of lightcurves across quarters, these scores can be more readily compared across quarters.
\begin{figure*}
    \centering
    \includegraphics[width=2\columnwidth]{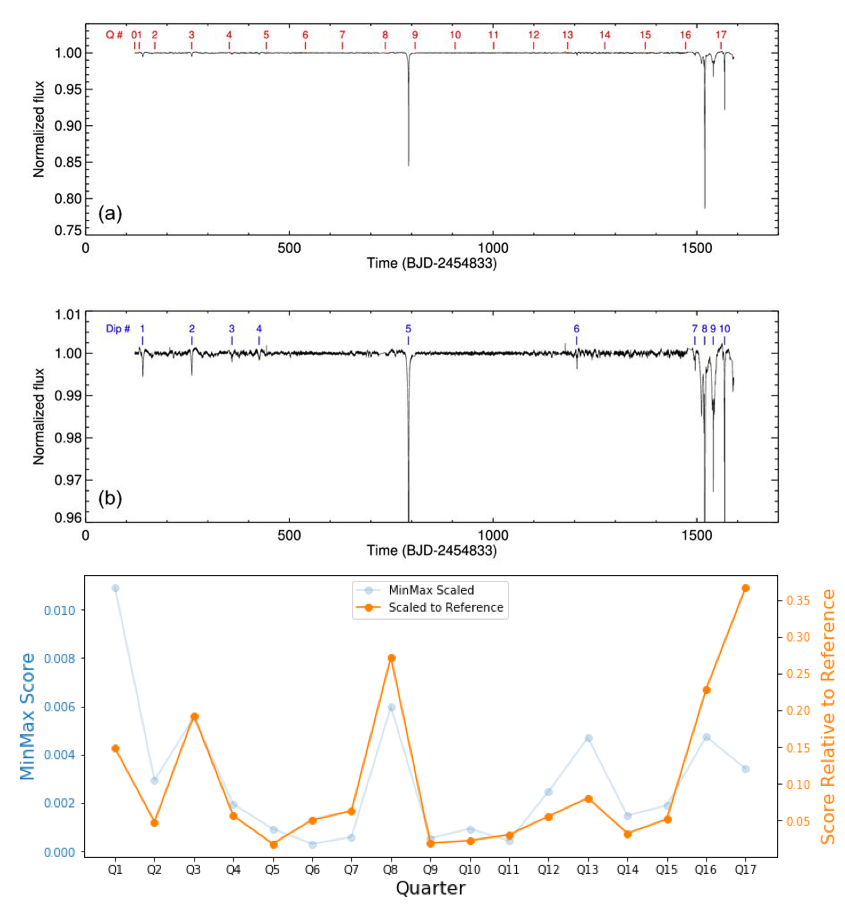}
    \caption{Light curve of Boyajian's star illustrating anomalous dips. \citep[Figure 1]{Boyajian2016}, scores for Boyajian's star by quarter. Using a common reference to scale scores facilitates comparison across quarters.}
    \label{fig:boyajian_lc_scores}
\end{figure*}

As an example that this produces more consistent score scaling across quarters, we look at the scores for Boyajian's star, a star which was discovered in the Kepler data by citizen scientists who noticed peculiar dips in several quarters \citep{Boyajian2016}. We show Boyajian star's lightcurve in the top 2 plots of Figure \ref{fig:boyajian_lc_scores} from \cite[Figure 1]{Boyajian2016}, which highlights the anomalous dips and demarcates the beginning and end of each quarter. 
In the bottom plot of Figure \ref{fig:boyajian_lc_scores}, we look at the scores from each quarter and see that they do not appear to be directly comparable as more extreme behaviour does not necessarily correspond to greater scores from the MinMax scaling. The ``Scaled to Reference'' line shows scores that are scaled to a reference, and these scores more closely match our intuition as the relative scores for each quarter mirror the more extreme behaviours better than the MinMax scaled scores. We provide these scores in Appendix \ref{app:alt_scores} Table \ref{tab:refscores}.

In addition to the table of scores for each object by quarter, we have produced a summary in Table \ref{tab:summary} to provide context for each object by identifier. This table provides the median rank and most outlying rank, plus a summary of information including coordinates, object type, and publications for the object available via the SIMBAD database. As a matter of convenience we also provide a table in Appendix \ref{app:files} containing the file names for each long cadence light curve file used in this work organized by Kepler ID and Quarter, which can be cross referenced to look up specific light curves via MAST.
\begin{landscape}
\begin{table} 
	\caption{Summary information for each object in the KIC. Rank information is from the k-average scores based on the full feature set. Position information, type, and bibliography are from the SIMBAD database.}
	\label{tab:summary}

	\begin{tabular}{cccccccc} 
	\hline
	KIC & 
	Median Rank &
	Minimum Rank &
	Quarter of Minimum &
	SIMBAD Object Type &
	RA &
	Dec &
	Biblio \\
    \hline
    757076 & 38099.0 & 8427.0 & Q14 & Star & 19 24 09.2898 & +36 35 53.121 & 2017ApJS..229...30M \\
    757099 & 2530.0 & 1563.0 & Q2 & Cepheid & 19 24 10.3300 & +36 35 37.602 & 2019MNRAS.484..834G|2016ApJ...829...23D|\dots \\
    757137 & 36285.0 & 13696.0 & Q11 & Eruptive* & 19 24 13.4198 & +36 33 35.724 & 2018ApJS..236...42Y|2016ApJ...829...23D \\
    757280 & 39413.0 & 8991.0 & Q10 &  &  &  &  \\
    757450 & 4610.0 & 2650.0 & Q7 & RotV* & 19 24 33.0185 & +36 34 38.477 & 2019AJ....158...59S|2019MNRAS.482.1379H|\dots \\
    \dots &\dots&\dots&\dots&\dots&\dots&\dots&\dots\\
	\hline
	\end{tabular}\\
Full, machine readable table available online at https://dx.doi.org/10.7910/DVN/H5QXUL.
\end{table}
\end{landscape}
\section{Conclusions}\label{sec:conclusions}
We have presented the results of outlier scoring on all long cadence light curves observed by the Kepler prime mission, providing scores for every lightcurve in the context of each quarter, as well as alternative scores scaled relative to an artificial reference to facilitate comparisons of scores by object across quarters.
We have shown that choice of neighbor, averaging over neighbors, and sampling for scoring minimally affect outlier scores. Furthermore, we've shown that scoring to high-k neighbors and sampled subsets become invariant to higher neighbors. The most dramatic changes in score come from scoring to dimensionality reductions and removing features, but even where scores can vary, the most outlying points identified from each approach overlap strongly with one another.
We have also shown using the elbow heuristic from DBSCAN that an overwhelming majority of the Kepler lightcurves outlier scores are negligible regardless of scaling or scoring methodology. 
It would seem that little is gained from additional granularity by faithfully calculating the distance to the exact nearest neighbors or even using PCA to mitigate correlation between features. Instead, optimizing computing performance to enable efficient application to additional and larger datasets to quickly identify and prioritize the most interesting data can take precedence.

This work establishes a method for identifying and prioritizing outliers in astronomical time domain surveys that may be applied to astronomical time domain surveys. The most immediate progression of this work is to scour the identified outliers and determine appropriate follow-up for each. Many of the most outlying points are data artifacts, which can be flagged for removal or reprocessing.
Pursuant to this work, the best basis for improvement for anomaly detection via a kNN approach will be better defining, selecting, and scaling features.
Another natural extension of this work is to improve upon the outlier scores and profile to generate probability estimates for identified outliers based on score distributions for known classes. This could better guide the use of generated scores. For example, clarifying if an outlier has a significant likelihood of being an exotic form of known phenomena, or if it is wholly unique.
We have only applied this methodology to the Kepler data so far, which has exceptionally well sampled and precise lightcurves. For application to future datasets, like LSST, we need to evaluate the performance on sparser datasets. Beyond archival data, this work should be applied to ongoing surveys to identify promising targets as well as catch pipeline issues early.

\section*{Acknowledgements}
DG thanks the Illinois Space Grant Consortium for supporting this work and the LSSTC Data Science Fellowship Program, his time as a Fellow has benefited this work. This research has made use of the SIMBAD database, operated at CDS, Strasbourg, France.

\section*{Data Availability}
The data underlying this article are available in Kepler Outlier Scores, at https://dx.doi.org/10.7910/DVN/H5QXUL.




\bibliographystyle{mnras}
\bibliography{references} 


\appendix
\section{Lightcurves}
\label{appendix:lightcurves}
In this appendix we provide lightcurves for some of the most outlying objects identified by our work. The lightcurves were obtained using the Observations module of Astroquery, and have been normalized in the same way as for feature calculation by dividing the flux values by the median flux. The lightcurves we present here are those that we used for feature calculation and scoring. In some cases where data artifacts led to negative fluxes, we have come to realize that the median flux value can be negative and flip the entire lightcurve.

\begin{figure*}
    \centering
    \includegraphics[width=1.8\columnwidth]{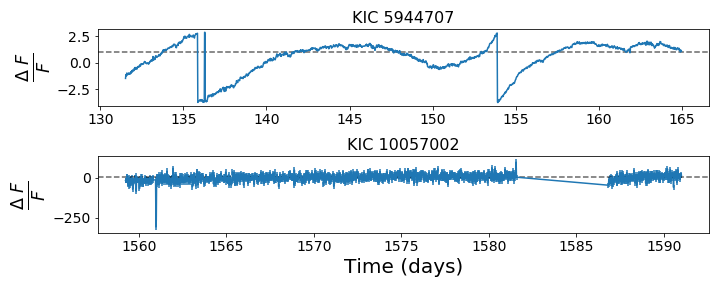}
    \caption{The most outlying lightcurves from the quarters 1 and 17 which are much shorter than the other quarters. These are the top outliers from the \krange{4}{14} scores based on the full feature set without sampling. For each lightcurve, the flux is normalized by its median flux value. These appear to be data artifacts with negative flux values. A dashed line has been placed at the normalized flux value of 1 as a visual aid.}
    \label{fig:top_outs_1_17}
\end{figure*}
\begin{figure*}
    \centering
    \includegraphics[width=1.8\columnwidth]{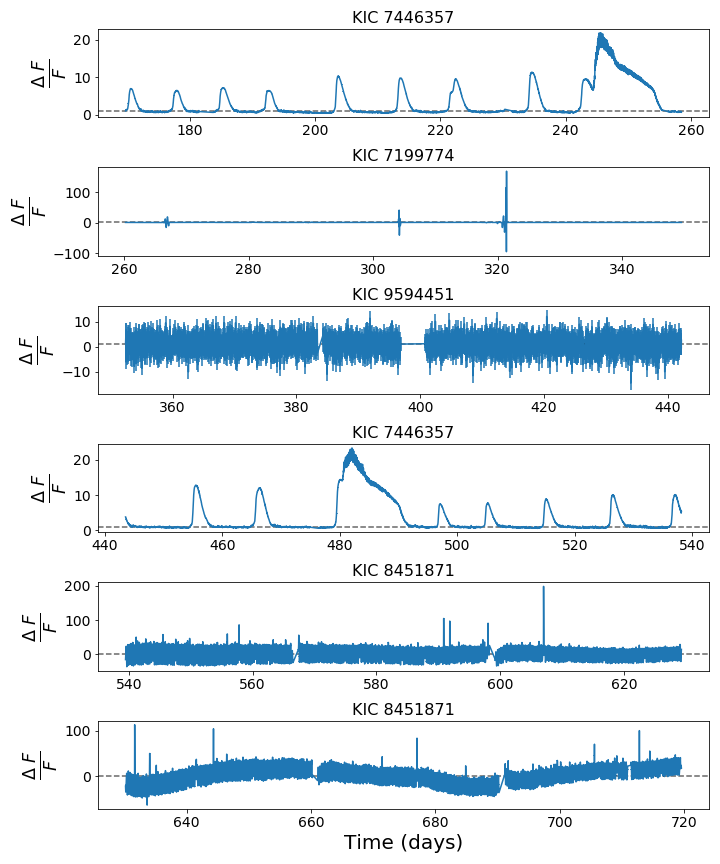}
    \caption{The most outlying lightcurves from quarters 2 to 7 from the \krange{4}{14} scores based on the full feature set without sampling. For each lightcurve, the flux is normalized by its median flux value. There is a mix between astrophysical anomalies like rare SU Uma Cataclysmic Variables and different data artifacts with negative flux values. A dashed line has been placed at the normalized flux value of 1 as a visual aid.}
    \label{fig:top_outs_2_7}
\end{figure*}
\begin{figure*}
    \centering
    \includegraphics[width=1.8\columnwidth]{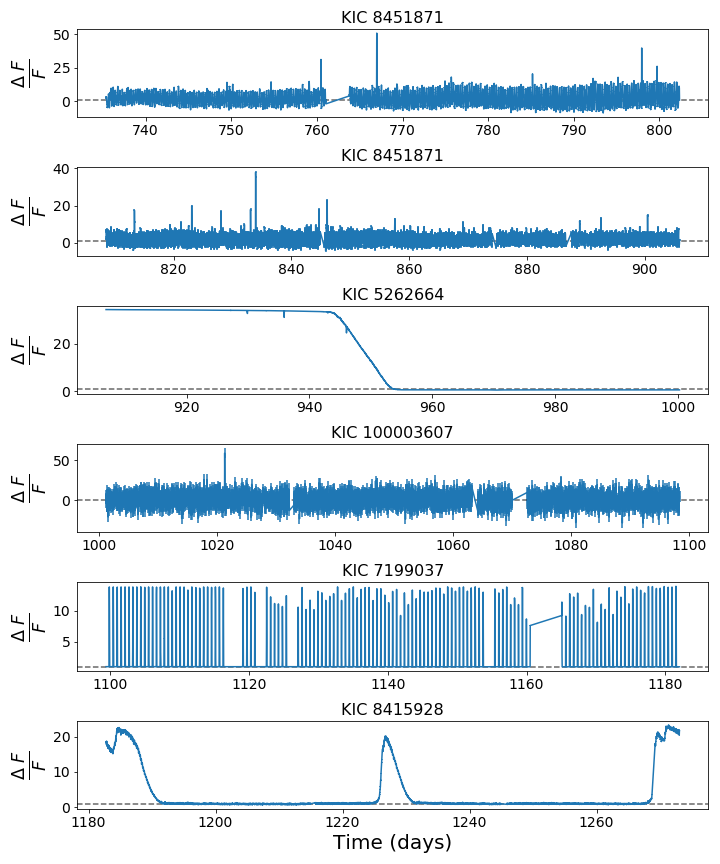}
    \caption{The most outlying lightcurves from quarters 8 to 13 from the \krange{4}{14} scores based on the full feature set without sampling. For each lightcurve, the flux is normalized by its median flux value. There is a mix between astrophysical anomalies like rare SU Uma Cataclysmic Variables and Eruptive stars, and different data artifacts, some of which with negative flux values. A dashed line has been placed at the normalized flux value of 1 as a visual aid.}
    \label{fig:top_outs_8_13}
\end{figure*}
\begin{figure*}
    \centering
    \includegraphics[width=1.8\columnwidth]{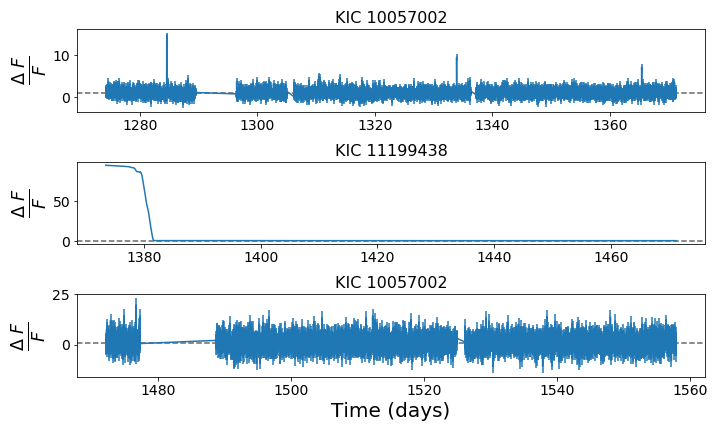}
    \caption{The most outlying lightcurves from quarters 14 to 16 from the \krange{4}{14} scores based on the full feature set without sampling. For each lightcurve, the flux is normalized by its median flux value. These appear to be different data artifacts, some of which with negative flux values. A dashed line has been placed at the normalized flux value of 1 as a visual aid.}
    \label{fig:top_outs_14_16}
\end{figure*}

\begin{figure*}
    \includegraphics[width=2\columnwidth]{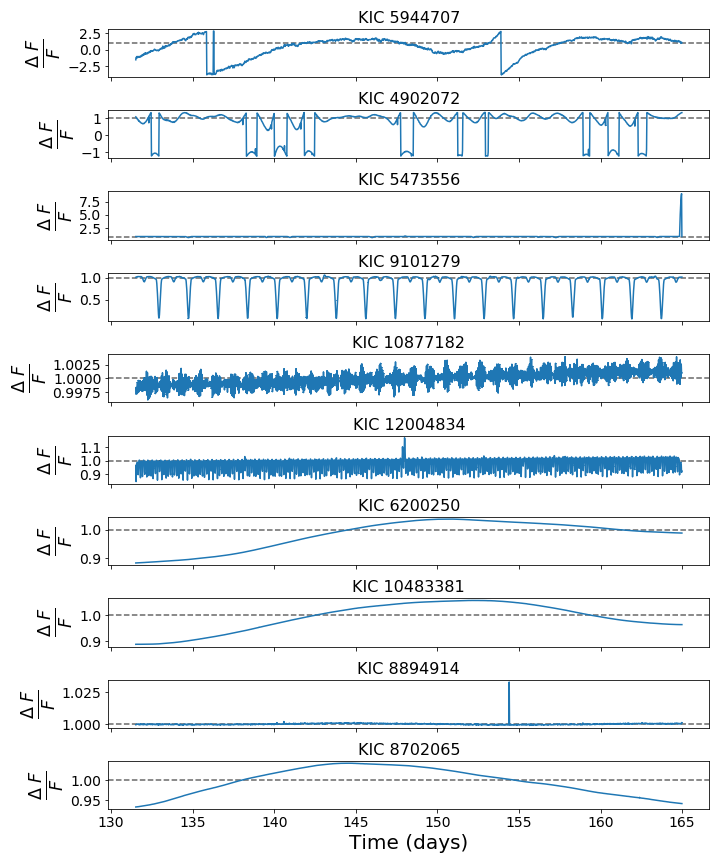}
    \caption{Top 10 outliers in Quarter 1 using the \krange{4}{14} scores based on the full feature set without sampling. Flux values have been normalized by the median flux value for each lightcurve and negative flux values reflect the data as available. A dashed line has been placed at the normalized flux value of 1 as a visual aid.}
    \label{fig:Q1topten}
\end{figure*}
\begin{figure*}
    \includegraphics[width=2\columnwidth]{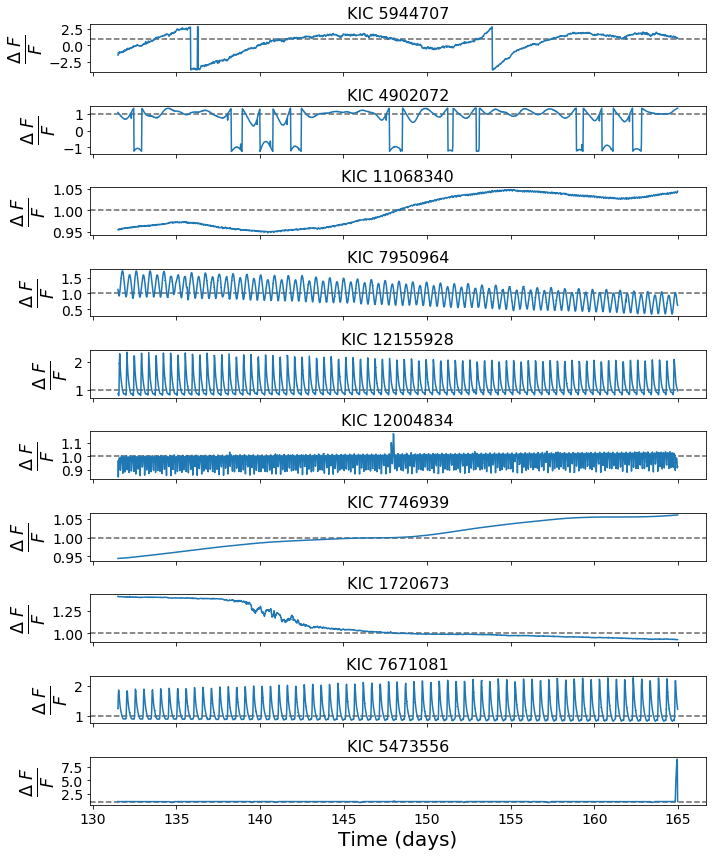}
    \caption{Top 10 outliers in Quarter 1 using the \krange{4}{14} scores based on the full feature set minus the flatness and roundness ratios. Flux values have been normalized by the median flux value for each lightcurve and negative flux values reflect the data as available. A dashed line has been placed at the normalized flux value of 1 as a visual aid.}
    \label{fig:Q1topten_ftsub}
\end{figure*}
\section{Kepler Files}
\label{app:files}
We provide a full machine readable table of the lightcurve file names we use in this work, accessible via MAST. This is provided for convenience and is organized by Kepler ID and quarter. 
\begin{table*} 
	\centering
	\caption{Full machine readable table of file names for long cadence lightcurves from MAST.}
	\label{tab:filenames}

	\begin{tabular}{ccccc}
    \hline
KIC & Q1 & Q2 & Q3 & ...\\

\hline
757076 & kplr000757076-2009166043257\_llc.fits & kplr000757076-2009259160929\_llc.fits & kplr000757076-2009350155506\_llc.fits & ...\\
757099 & kplr000757099-2009166043257\_llc.fits & kplr000757099-2009259160929\_llc.fits & kplr000757099-2009350155506\_llc.fits & ...\\
757137 & kplr000757137-2009166043257\_llc.fits & kplr000757137-2009259160929\_llc.fits & kplr000757137-2009350155506\_llc.fits & ...\\
757280 & kplr000757280-2009166043257\_llc.fits & kplr000757280-2009259160929\_llc.fits & kplr000757280-2009350155506\_llc.fits & ...\\
757450 & kplr000757450-2009166043257\_llc.fits & kplr000757450-2009259160929\_llc.fits & kplr000757450-2009350155506\_llc.fits & ...\\
... & ... & ... & ... & ...\\
		\hline
	\end{tabular}\\
Full, machine readable table available at https://dx.doi.org/10.7910/DVN/H5QXUL.
\end{table*}

\section{Alternative Scores}
\label{app:alt_scores}
We include here an alternate table of scores scaled with respect to a reference which facilitates comparisons of scores across quarters. Each column is a linear transformation of the corresponding column of Table \ref{tab:minmaxscores} with differing weights.
\begin{table*} 
	\centering
	\caption{Outlier scores scaled with respect to an artificial reference source. These scores are exact k=1, sampled 10$\times$1,000.}
	\label{tab:refscores}

	\begin{tabular}{lrrrrrrrrrrrrrrrrr} 
    \hline
 & Q1 & Q2 & Q3 & Q4 & Q5 & Q6 & Q7 & Q8 & Q9 & Q10 & Q11 & Q12 & Q13 & Q14 & Q15 & Q16 & Q17 \\
KIC & $10^{-1}$ & $10^{-1}$ & $10^{-1}$ & $10^{-1}$ & $10^{-1}$ & $10^{-1}$ & $10^{-1}$ & $10^{-1}$ & $10^{-1}$ & $10^{-1}$ & $10^{-1}$ & $10^{-1}$ & $10^{-1}$ & $10^{-1}$ & $10^{-1}$ & $10^{-1}$ & $10^{-1}$ \\
\hline
757076 & 0.15 & 0.20 & 0.23 & 0.15 & 0.14 & 0.63 & 0.41 & 0.12 & 0.27 & 0.30 & 0.72 & 0.15 & 0.45 & 0.55 & 0.27 & 0.29 & 0.55 \\
757099 & 0.82 & 2.38 & 1.40 & 1.39 & 1.02 & 4.20 & 2.57 & 1.22 & 1.18 & 1.26 & 2.75 & 1.61 & 1.44 & 1.90 & 1.66 & 1.43 & 2.60 \\
757137 & 0.17 & 0.23 & 0.19 & 0.42 & 0.23 & 0.74 & 0.92 & 0.36 & 0.24 & 0.27 & 0.88 & 0.37 & 0.27 & 0.23 & 0.27 & 0.62 & 0.80 \\
757280 & 0.23 & 0.23 & 0.34 & 0.26 & 0.21 & 0.87 & 0.76 & 0.39 & 0.47 & 0.62 & 0.40 & 0.19 & 0.15 & 0.17 & 0.22 & 0.49 & 0.46 \\
757450 & 0.66 & 0.64 & 0.83 & 0.62 & 0.86 & 3.26 & 2.49 & 1.01 & 1.02 & 0.79 & 1.59 & 0.80 & 0.93 & 0.79 & 0.79 & 0.61 & 1.42 \\
... & ... & ... & ... & ... & ... & ... & ... & ... & ... & ... & ... & ... & ... & ... & ... & ... & ... \\

	\hline
	\end{tabular}
Full, machine readable table available at online at  https://dx.doi.org/10.7910/DVN/H5QXUL.
\end{table*}

\bsp	
\label{lastpage}
\end{document}